\documentclass[Proceedings]{ascelike}
\usepackage{graphicx}
\usepackage{bm}
\usepackage{amsmath}
\usepackage{amsfonts}
\usepackage{amssymb}
\usepackage{amsbsy}
\usepackage{times}
\usepackage{color}
\NameTag{Takada, \today}
\begin{document}
%
\title{Drag Law of Two Dimensional Granular Fluids}
\author{
Satoshi Takada%
%
\thanks{
Yukawa Institute for Theoretical Physics,
Kyoto University,
Kyoto 606-8502, Japan. E-mail: stakada@yukawa.kyoto-u.ac.jp.}
%
%
 and
 Hisao Hayakawa%
 \thanks{Yukawa Institute for Theoretical Physics,
Kyoto University,
Kyoto 606-8502, Japan. E-mail: hisao@yukawa.kyoto-u.ac.jp.}
}
\maketitle
\begin{abstract}
The drag force law acting on a moving circular disk in a two-dimensional granular medium 
is analyzed based on the discrete element method (DEM).
It is remarkable that the drag force on the moving disk in moderate dense and pure two-dimensional granular medium 
can be well reproduced by a perfect fluid with separation from the surface of the tracer.
A yield force, being independent of the moving speed of the disk, appears
if a dry friction between the granular disks and the bottom plate exists.
The perfect fluidity is violated in this case.
The yield force and the drag force diverge at the jamming point.
\end{abstract}
\KeyWords{drag law, DEM, perfect fluidity, jamming transition.}
\section{Introduction}
The drag law of a moving object in a medium is one of the most fundamental characterizations of rheology. 
The drag force on an object surrounded by a viscous fluid is proportional to the moving speed, 
whose dependence is known as Stokes law, if the moving speed is low.  
The drag is proportional to a fractional power of the moving speed, 
when the speed is faster.
Then the power exponent reaches 2 in the high speed limit \cite{Lamb:1945,Batchelor:1967}.

The drag force on a moving object in a granular medium is completely
different from that in the viscous fluid. 
Because granular materials behave as unusual solids and liquids \cite{Jarger:1996},
 the drag force is thought to be a complex combination of the force chains and fluid contribution.
So far, some previous experiments reported 
that the drag force has 
non-trivial depth-dependence on the drag of cylinders \cite{Albert:1999,Hill:2005,Guillard:2013} and
logarithmic dependence on the rotating frequency in a two-dimensional geometry and in a biaxial rotating cylinder \cite{Geng:2004,Geng:2005,Reddy:2011}.

Most of these previous studies indicated the existence of two drag terms: 
one is independent of the moving speed, and another depends on that.
Recently, Takehara and her coworkers have performed a series of experiments 
to clarify the drag law acting on a circular disk 
in one layer granular medium by controlling the pulling speed $V$ of the disk \cite{Takehara:2010,Takehara:2014}. 
They experimentally found that 
the drag force in a granular medium satisfies
\begin{equation}
F_{\rm drag}=F_0(\phi)+\alpha(\phi) V^2,\label{eq:okumura}
\end{equation}
where both $\alpha(\phi)$ and $F_0(\phi)$ are proportional to $(\phi_{\rm c}-\phi)^{-1/2}$
with the area fraction $\phi$ and the jamming fraction $\phi_{\rm c}$ \cite{Takehara:2014}.
It should be noted that 
Eq.\ (\ref{eq:okumura}) may be only valid for the moving object at relatively high speed 
as mentioned in their paper \cite{Takehara:2014}, 
though the expression itself can be extrapolated to $V\to 0$. 
Some experiments on the impact of a hard projectile into granular beds 
also supported Eq.\ (\ref{eq:okumura}), 
though the density dependencies of the two terms are different
\cite{Katsuragi:2007,Katsuragi:2013,Clark:2012,Seguin:2009}.
Although a previous study for three-dimensional simulation 
reported that the drag force is proportional to $V$ \cite{Hilton:2013}
and the other studies \cite{Chehata:2003,Wassgren:2003,Geng:2004,Geng:2005,Bharadwaj:2006} 
suggested that the drag force logarithmically depends on $V$. Nevertheless,
the drag force proportional to $V^2$ is natural for granular systems, 
which is consistent with Bagnold's scaling \cite{Bagnold:1954} 
and the direct impulse of granular particles hitting a moving object.
The existence of the force being independent of the moving speed 
does not correspond to the drag law in a viscous fluid.
To verify the validity of Eq.\ (\ref{eq:okumura}) and 
understand the mechanism to appear Eq.\ (\ref{eq:okumura})
are the central issues of this paper.

Takehara and Okumura (2014) suggested that the existence of $F_0(\phi)$ is related to the jamming transition \cite{Liu:1998,O'Hern:2002,O'Hern:2003,Olsson:2007,Hatano:2008,Otsuki:2009b,Otsuki:2009a,Otsuki:2011,Tighe:2010,Nordstrom:2010}.
It is, however, obvious that the jamming transition is unrelated to the existence of $F_0(\phi)$, 
because $F_0(\phi)$ still exists far below the jamming density.
The measurement of the drag force is relevant to know the viscosity in a viscous fluid.
The viscosity of the granular fluid, however, obtained from the Couette flow 
is much larger than that obtained from the drag experiment.
Indeed, an experiment for a granular jet \cite{Chen:2007} as well as simulations \cite{Ellowitz:2013,Mueller:2014} suggested that the granular fluid can be approximately represented by a perfect fluid,
though there exist counter arguments \cite{Sano:2012,Sano:2013}. 
Therefore, another purpose of this paper is to resolve the current confusing situation 
on the rheology of granular fluids.
\begin{figure}
	\begin{center}
	\includegraphics[width=40mm]{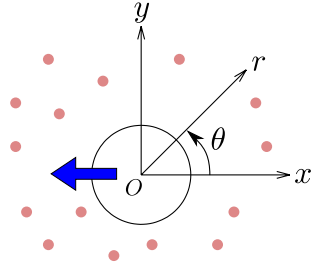}
	\end{center}
	\caption{(Color online) A schematic picture of our setup. 
			We choose the pulling direction of the tracer as negative $x$-direction.
			We also introduce the polar coordinates ($r$, $\theta$).
			Here, the arrow represents the moving direction of the tracer.}
	\label{fig:coordinate}
\end{figure}

\section{Model}
In this paper, we perform two-dimensional simulations in terms of the discrete element method (DEM) \cite{Cundall:1979}
for a moving disk (the diameter $D$, the mass $M$, the position $\bm{R}$, and the velocity $\bm{V}$)
surrounded by granular particles 
(the diameter $d_i$, the mass $m_i$, the position $\bm{r}_i$, and the velocity $\bm{v}_i\equiv \dot{\bm{r}}_i$ for $i$-th grains and the number of grains is $N$)
with or without the influence of dry friction characterized by Coulombic friction constant $\mu$
between the bottom plate and the granular disks (Fig.\ \ref{fig:coordinate}).
Here, we consider both cases with and without the rotation of the disks and the tangential contacting forces.
For the frictionless case, 
the equations of motion of the tracer which is the moving object in this study and the granular particle $i$ are expressed as
\begin{equation}
\begin{cases}
M \ddot{\bm{R}}=\bm{F}_{\rm ex}+\bm{F}_{\rm int}(\bm{R})-\mu Mg \hat{\bm{V}},\\
m_i \ddot{\bm{r}}_i= \bm{F}_{\rm int}(\bm{r}_i)-\mu m_ig \hat{\bm{v}}_i,
\end{cases}\label{eq:EOM}
\end{equation}
where $\bm{F}_{\rm int}$ represents the interaction between grains satisfying 
$\bm{F}_{\rm int}(\bm{r}_i)={\sum'}_{j=0}^{N}\bm{f}^n_{ij}$ 
with $\bm{f}^n_{ij}= \Theta(d_{ij}-r_{ij})\{ k_n(d_{ij}-r_{ij})\hat{\bm{r}}_{ij}- \eta_n \dot{\bm{r}}_{ij}\cdot\hat{\bm{r}}_{ij}\}$,
and $\hat{\bm{V}}= \bm{V}/|\bm{V}|$ and $\hat{\bm{v}}_i=\bm{v}_i/|\bm{v}_i|$
are unit vectors parallel to $\bm{V}$ and $\bm{v}_i$, respectively.
Here, ${\sum}'$ denotes the summation under the condition $j\ne i$, and
$\Theta(x)$ is the step function, i.e.  $\Theta(x)=1$ for $x\ge 0$ and $\Theta(x)=0$ for otherwise.
We characterize the tracer by $i=0$, $d_0=D$, and $\bm{r}_0=\bm{R}$.
We also use $d_{ij}=(d_i+d_j)/2$,
the relative velocity $\dot{\bm{r}}_{ij}$ between $i$ and $j$ grains, 
and $\bm{\hat{r}}_{ij}=(\bm{r}_i-\bm{r}_j)/| \bm{r}_i-\bm{r}_j|$ 
with the spring constant $k_n$ and the viscous parameter $\eta_n$ in the normal direction.
It should be noted that Hertzian contact force in a two-dimensional system can be written 
as a term proportional to the compression with a logarithmic correction \cite{Johnson:1985,Gerl:1999,Hayakawa:2002}.
In this paper, for simplicity, we adopt the linear spring model to represent the elastic force 
between contacting particles.

For the frictional case, Eq.\ (\ref{eq:EOM}) still can be used with the replacement of the contact force by 
$\bm{F}_{\rm int}(\bm{r}_i)={\sum'}_{j=0}^{N} (\bm{f}^n_{ij} + \bm{f}^t_{ij})$,
where the tangential force $\bm{f}^t_{ij}$ is given by

\begin{align}
\bm{f}^t_{ij}= \Theta(d_{ij}-r_{ij}) \min (\mu_s|\bm{f}^n_{ij}|, |\tilde{\bm{f}}^t_{ij}| )\bm{t}_{ij},
\end{align}
where we adopt $\mu_s=0.2$ for Coulombic friction constant between grains, $\tilde{\bm{f}}^t_{ij}=-k_t \bm{\xi}_{ij}-\eta_t \bm{v}_{t,ij}$,
and $\bm{t}_{ij}=\tilde{\bm{f}}^t_{ij}/|\tilde{\bm{f}}^t_{ij}|$.
Here, 
$\bm{v}_{t,ij}=\dot{\bm{r}}_{ij}+\hat{\bm{r}}_{ij}\times (d_i\bm{\omega}_i+d_j \bm{\omega}_j)
-(\dot{\bm{r}}_{ij}\cdot \hat{\bm{r}}_{ij})\hat{\bm{r}}_{ij}$ with the angular velocity $\bm{\omega}_i$ 
of $i$-th particle,
$\bm{\xi}_{ij}$ 
is the tangential overlap vector between $i$-th and $j$-th particles
defined by
$\bm{\xi}_{ij}=\int_{t_0}^t dt^\prime \bm{v}_{t,ij}(t^\prime)$
with the time $t_0$ of first contact of the particles,
and $k_t$ and $\eta_t$ are the spring constant and the viscous parameter 
in the tangential direction, respectively.
We have introduced the function ${\rm min}(a,b)$ to select smaller one from $a$ and $b$.
In addition, an equation of motion for the rotation is given by
\begin{align}
I_i \dot{\bm{\omega}}_i &= \bm{T}_i - \zeta_\omega \bm{\omega}_i,
\end{align}
where $\dot{\bm{\omega}}_i$ is the angular acceleration of $i$-th particle,
$I_i=m_i d_i^2/8$ is the moment of inertia of $i$-th particle, 
$\bm{T}_i=\sum_{j=0}^{\prime N}\Theta(d_{ij}-r_{ij}) \{ \bm{r}_i \times (\bm{f}^n_{ij} + \bm{f}^t_{ij}) \}$ is the torque of $i$-th particle, 
and $\zeta_\omega$ is the rolling friction constant \cite{Suyama:2008}.
We assume that the mass density of each grain is identical and the system contains an equal number of two types of grains characterized by the diameters $d$ and $1.4d$ to avoid the crystallization.
It should be noted that the driving force $\bm{F}_{\rm ex}=-F_{\rm ex} \hat{\bm{e}}_x$ 
directly acts on the tracer i.e. moving disk, where $\hat{\bm{e}}_x$ is the unit vector in $x$-direction.
The system reaches a steady state by the balance between $\bm{F}_{\rm ex}$ and the other forces, $\bm{F}_{\rm int}(\bm{R})$ and the dry friction force $-\mu Mg\hat{\bm{V}}$ between the tracer and the bottom plate.
Thus, we obtain a steady motion of the moving disk to negative $x$-direction from the simulation of Eq.\ (\ref{eq:EOM}).
We adopt the values of parameters
$\eta_n=0.75\sqrt{mk_n}$ corresponding to the restitution constant $e=0.92$,
$\zeta_\omega=447md\sqrt{m/k_n}$
and the time increment $\Delta t=0.002\sqrt{m/k_n}$, 
where $m$ is the mass of the grain of the diameter $d$.
The value of $\zeta_\omega$ is equal to $\zeta_\omega=20md\sqrt{d/g}$ 
in terms of the gravitational acceleration $g$ instead of $k_n$. 
We have checked that the motion of the tracer is almost same in the range $10md\sqrt{d/g}\le \zeta_\omega \le 50md\sqrt{d/g}$ when we perform the simulations using the identical initial condition.
The values of $k_t$ and $\eta_t$ are chosen as $k_t=(2/7)k_n$ and $\eta_t=(2/7)\eta_n$, respectively \cite{Thompson:1991} and $k_n$ is chosen as $k_n=500mg/d$ with the gravitational acceleration $g$.
For dry friction between the grains and the bottom plate, we examine $\mu=0.001$ and $\mu=0$.
The reason why we adopt such a small value for $\mu$ is 
that it is difficult to obtain a steady motion for a wide range of the external force for larger $\mu$.
We adopt the velocity Verlet algorithm for the time integration of the equations of motion.
 The system size is basically fixed to be $210d \times 105d$.
 As a result, the number of grains $N$ depends on the area fraction, though a typical value is $N\sim 2\times 10^4$.
 The grains are located at random without any motion and overlap at time $t=0$.
 For simplicity, we adopt the periodic boundary condition for $x$-direction and 
 we assume that the boundary in $y$-direction is composed of particles whose mass and curvature are infinite with the spring constants $k_n$, $k_t$ and the viscous parameters $\eta_n$, $\eta_t$, respectively.
 We have two control parameters, the area fraction $\phi$ where the tracer is not included and the external force $F_{\rm ex}$.
\section{Results}
\begin{figure}[htbp]
	\begin{center}
		\includegraphics[width=120mm]{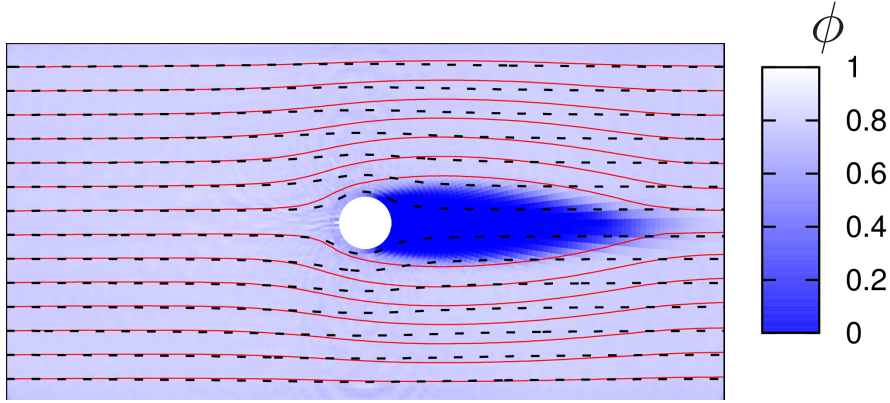}
		\caption{(Color online) The density profile (color scale) and the streamlines (the red solid lines) 
				obtained from our DEM of frictional grains for $\phi=0.76$, $F_{\rm ex}=0.2k_nd$, and 
				the streamlines of the perfect fluid (the black dashed lines), where we have used $D=10d$. 
				Here the flow direction is from left to right in the frame that the tracer is stationary.}
		\label{fig:stream_line}
	\end{center}
\end{figure}
\begin{figure}[htbp]
	\begin{center}
		\includegraphics[width=145mm]{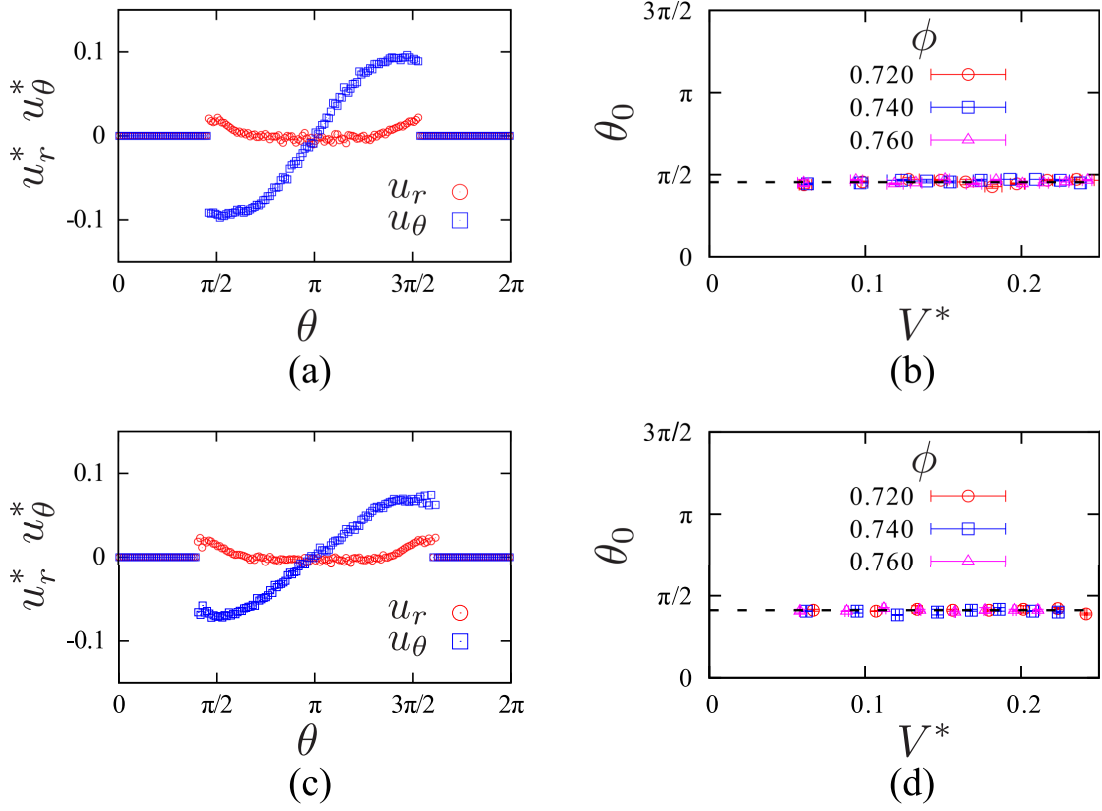}
	\end{center}
	\caption{(Color online) (a) The velocity field around the tracer against 
			$\theta$ for $F_{\rm ex}=0.2k_nd$ and $\phi=0.76$ for the frictionless disks 
			where we plot the radial component $u_r$ (the red open circles) 
			and the polar component $u_\theta$ (the blue open squares). 
			(b) The velocity dependence on the separation angle $\theta_0$ for the frictionless disks,
			where the separation angle is defined by the angle where $u_\theta$ deviates from a sinusoidal function.
			(c) The velocity field against $\theta$ for the frictional disks.
			(d) The velocity dependence on $\theta_0$ for the frictional disks.
			The dashed line is the average over the results.
			Here, we have introduced dimensionless quantities $u_r^*\equiv u_r \sqrt{m/k_n}/d$,
			$u^*_\theta=u_\theta \sqrt{m/k_n}/d$ and $V^*=V \sqrt{m/k_n}/d$.}
	\label{fig:boundary}
\end{figure}
\begin{figure}[htbp]
	\begin{center}
		\includegraphics[width=145mm]{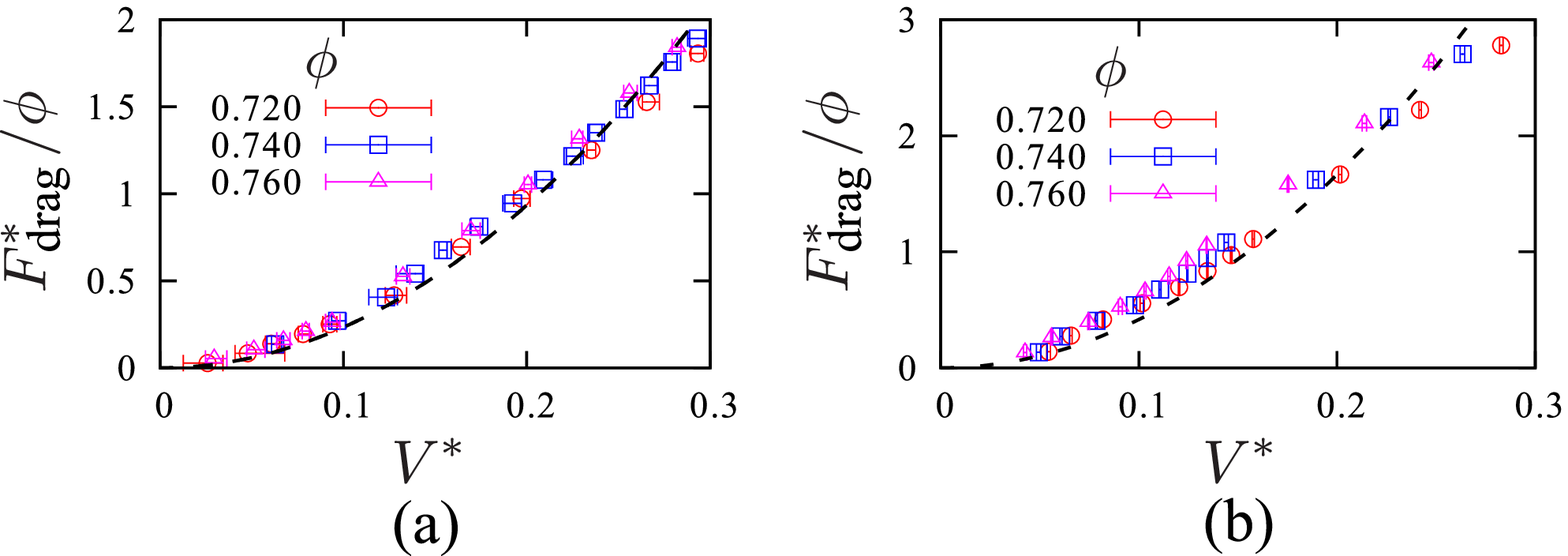}
	\end{center}
	\caption{(Color online) The dimensionless drag force 
			$F^*_{\rm drag}\equiv F_{\rm drag}/k_nd$ 
			divided by $\phi$ against  $V^*$
			for (a) the frictionless disks and (b) the frictional disks, respectively, for three densities:  
			$\phi=0.72$ (red open circles), 
			$\phi=0.74$ (blue open squares), and
			$\phi=0.76$ (pink open triangles),
			with $F^*_0\equiv F_0/k_nd$.
			The dashed lines are theoretical ones obtained from the perfect fluid 
			where the separation angles are, respectively, determined 
			by Figs.\ \ref{fig:boundary}(b) and \ref{fig:boundary}(d) for frictionless and frictional grains.}
	\label{fig:density}
\end{figure}

\subsection{System without dry friction}
First, we analyze a frictionless or a frictional system consisting of a large tracer disk and a collection of granular disks without the influence of the dry friction i.e. $\mu=0$ in Eq.\ (\ref{eq:EOM}).
Figure \ref{fig:stream_line} displays the density profile obtained from our DEM for $F_{\rm ex}=0.2k_nd$ and $\phi=0.76$,
 and the streamlines of both our DEM (red solid lines) and the perfect fluid (dashed lines) in the frame that the tracer is stationary.
 The streamlines are obtained by averaging over time during $100\sqrt{m/k_n}$ and 10 ensembles.
It is notable that the density is almost uniform except for the cavity right behind the tracer, where no grains exist in this region.
The streamlines obtained by the DEM are smooth and do not contain any vortex excitation.
It is remarkable that the streamlines of the DEM are well reproduced 
by those of the perfect fluid except for that in the cavity.
We stress that there is no contribution to the drag force from the cavity, 
because there are no grains colliding to the tracer there.

Figures \ref{fig:boundary}(a) and (c) represent the velocity fields around the tracer against the polar angle $\theta$ 
for $F_{\rm ex}=0.2k_nd$ and $\phi=0.76$ for frictionless grains and frictional grains, respectively. 
These figures clearly support that the radial component of the granular flow on the surface of the tracer is almost zero, and the polar component of the granular flow can be approximately represented by a sinusoidal function of $\theta$ for $\theta_0<\theta<2\pi-\theta_0$ and 0 for $-\theta_0<\theta <\theta_0$.
Here, the separation angle, $\theta_0$ (which is nearly equal to $80^\circ$ for frictionless disks and $70^\circ$ for frictional disks), is almost independent of $V$ and $\phi$ (Figs.\ \ref{fig:boundary}(b) and (d)).
If $\theta_0=\pi/2$, it can be interpreted as the result of direct impulses of colliding grains.
However, $\theta_0$ is a little smaller, as seen in Figs.\ \ref{fig:boundary}(b) and (d).
The mechanism to have smaller $\theta_0$ might be understood by the finite granular temperature effect.

Let us examine the drag force of the perfect fluid with the separation angle $\theta_0$.
It is well known that the pressure around a cylinder in the perfect fluid is given by
\begin{equation}
p=p_\infty+\frac{\rho}{2}V^2(1-4\cos^2\theta), 
\end{equation}
for an irrotational incompressible perfect fluid, 
where $\rho$ is the mass density of the granular fluid \cite{Batchelor:1967}.
Because the far field pressure $p_\infty$ is the impulse per unit cross line at the boundary, we adopt the expression $p_\infty= (1+e) \rho V^2$, where $e$ is the restitution constant.  
 From the integration of the pressure acting on the surface of the tracer, 
$F_{\rm drag}=-\int_{\theta_0}^{2\pi-\theta_0}(D/2)d\theta p\cos\theta$,
we obtain the drag force
\begin{equation}
 F_{\rm drag}=\left(\frac{3+2e}{2}-\frac{2}{3}\sin^2\theta_0\right) \sin\theta_0 D \rho V^2.\label{eq:drag_law}
\end{equation}
 Note that $F_{\rm drag}$ should be zero if there is no separation, $\theta_0=0$, in Eq.\ (\ref{eq:drag_law}).
The granular fluid, however, has a finite contribution even if the viscosity is zero 
because of the separation of the flow \cite{Southwell:1946}.
 The expression (\ref{eq:drag_law}) indicates that there is no yield force for pure two-dimensional cases,
 in contrast to previous experimental results \cite{Takehara:2010,Takehara:2014}.
It is astonishing that this simple formula (\ref{eq:drag_law}) well reproduces the result of our simulation for $\phi=0.72, 0.74$ and $0.76$ without any fitting parameter (Fig.\ \ref{fig:density}),
while the formula (\ref{eq:drag_law}) deviates from the simulation results when $V^\ast\equiv V\sqrt{m/k_n}/d$ is larger than 0.3.
This result is interesting because the perfect fluidity observed in granular jets or granular fluids \cite{Chen:2007,Ellowitz:2013,Mueller:2014,Blumenfeld:2010} is quantitatively verified in our setup, at least, for relatively slow flows and moderate dense granular medium.
We should note that our problem can be converted into a jet problem for a circular target, if we use the frame of the stationary tracer.
Because the flow has zero granular temperature at $t=0$, the excitation after the impact can be the origin of the viscosity.
However, if the flow is slow and the target is circular, the excitation of the temperature by collisions is quite small and, thus, the flow in our setup can keep the perfect fluidity.
We also note that Eq.\ (\ref{eq:drag_law}) is no longer valid in the vicinity of the jamming point.

\subsection{The role of dry friction}
\begin{figure}[htbp]
	\begin{center}
		\includegraphics[width=145mm]{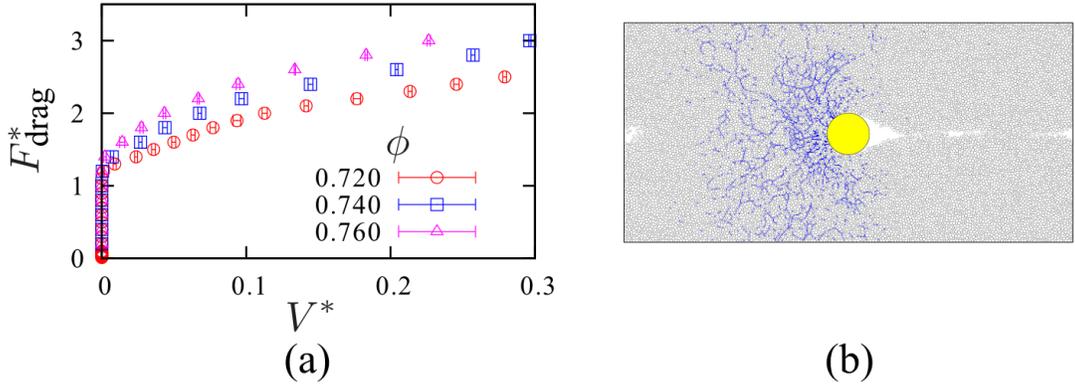}
	\end{center}
	\caption{(Color online) (a) The relationship between $F^*_{\rm drag}$ and $V^*$ 
			for frictionless disks in
			$\phi=0.72$ (red open circles), 
			$\phi=0.74$ (blue open squares) and 
			$\phi=0.76$ (pink open triangles).
			(b) A snapshot of the force chains (blue solid lines) for frictionless disks 
			at $\phi=0.82$, $F^*_{\rm drag}=3.0$,
			where line width represents the strength of the interaction between contacting particles.}
	\label{fig:density_mu}
\end{figure}

Now, let us consider the case of finite friction, i.e., $\mu \ne 0$ between the bottom plate and grains.
For $\mu\ne 0$, 
the gravitational acceleration $g$ can produce a new time scale $\sqrt{d/g}$.
Therefore, we expect that the yield force $F_0$ is finite for $\mu \ne 0$.
Figure \ref{fig:density_mu}(a) exhibits the result of our DEM for $\mu=10^{-3}$.
In this case, however, the perfect fluidity is violated.
This violation might be related to the existence of force chains as in Fig.\ \ref{fig:density_mu}(b).
Namely, the motion of grains is correlated with each other through the force chains.

\begin{figure}[htbp]
	\begin{center}
		\includegraphics[width=145mm]{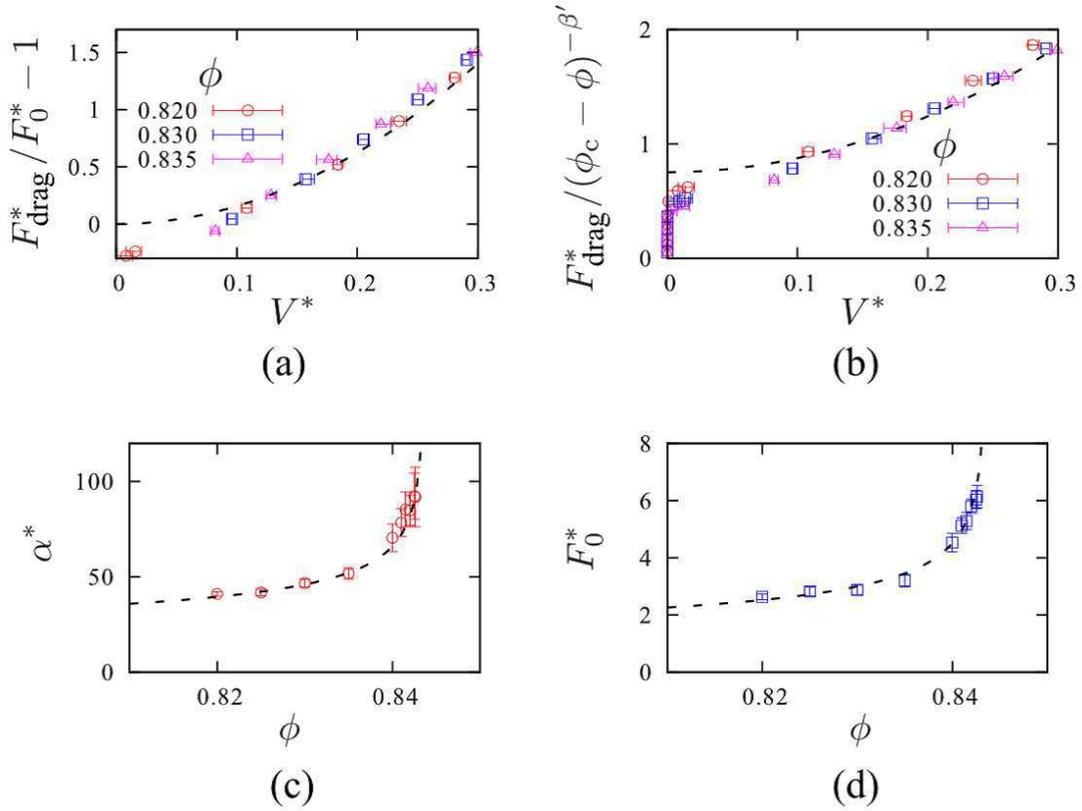}
	\end{center}
	\caption{(Color online) (a) The plot of $F^*_{\rm drag}$ 
			divided by $F^*_0$ against $V^*$ 
			for frictionless disks in
			$\phi=0.82$ (red open circles), 
			$\phi=0.83$ (blue open squares), and 
			$\phi=0.835$ (pink open triangles).
			The dashed line is a fitting curve in the range $0.1\le V^*\le 0.3$ 
			using Eq.\ (\ref{eq:okumura}).
			(b) The plot of $F^*_{\rm drag}$ 
			divided by $(\phi_{\rm c}-\phi)^{-\beta^\prime}$ 
			against $V^*$ for $\phi=0.82$, $0.83$, and $0.835$
			with $\phi_{\rm c}=0.8437$ and $\beta^\prime=0.312\pm0.027$.
			The dashed line is a fitting curve in the range $0.1\le V^*\le 0.3$ 
			using Eq.\ (\ref{eq:okumura}).
			(c) The density dependence of $\alpha$, where dashed line is given by $\alpha(\phi)\sim (\phi_{\rm c}-\phi)^{-\beta}$ 
			with $\beta=0.277\pm0.028$.
			(d) The density dependence of $F_0$, 
			where dashed line is given by $F_0(\phi)\sim (\phi_{\rm c}-\phi)^{-\beta^\prime}$.}
	\label{fig:F0_alpha}
\end{figure}

As the density increases, the drag force increases.
Then it diverges at the jamming point.
It seems that Eq.\ (\ref{eq:okumura}) is still valid, at least, for $0.1\le V^*\le 0.3$.
In other words, we cannot fit the data by neither a logarithmic function nor a linear function of $V$ for the drag.
We determine the coefficient $\alpha$ in Eq.\ (\ref{eq:okumura}) by fitting the data in the range $0.1\le V^*\le0.3$, 
where $F_0$ is estimated by extrapolating to $V\to 0$.
Figures \ref{fig:F0_alpha}(a) and (b) show that 
$\alpha$ and $F_0$ have the almost identical dependencies on the density as
\begin{align}
\alpha(\phi) \sim& (\phi_{\rm c}-\phi)^{-\beta},\label{eq:alpha}\\
F_0(\phi) \sim& (\phi_{\rm c}-\phi)^{-\beta^\prime},\label{eq:F0}
\end{align}
near the jamming point $\phi_{\rm c}$
where $\phi_{\rm c}=0.8437$, $\beta=0.277\pm0.028$, and $\beta^\prime=0.312\pm0.027$, respectively (see Figs.\ \ref{fig:F0_alpha}(c) and (d)).
It should be noted that the drag law deviates from the quadratic form for $V^*\le 0.1$.
Similar behavior is also observed in a previous experiment \cite{Okumura:2014}.
We stress that the jamming point $\phi_{\rm c}$ in Eqs.\ (\ref{eq:alpha}) and (\ref{eq:F0})
which is a little larger than the previous estimations \cite{Otsuki:2009b,Otsuki:2012,Takehara:2014}.
We also note that the results cannot be represented by simple power laws as in Eqs.\ (\ref{eq:alpha}) and (\ref{eq:F0}) if we choose smaller $\phi_{\rm c}$ reported in the previous studies.
The exponents of $\beta$ and $\beta^\prime$ in Eqs.\ (\ref{eq:alpha}) and (\ref{eq:F0}) are a little smaller than those by Takehara and Okumura (2014),
though $\beta$ is nearly equal to $\beta^\prime$.
Therefore, we conclude that our simulation near the jamming point is qualitatively similar to that observed by Takehara and Okumura (2014) but quantitative agreement is poor, which might be from the softness of the particles in our simulation.

\section{Conclusion}
In this paper, we performed two-dimensional DEMs to study the drag force acting on the tracer for both frictionless and frictional granular disks with or without the influence of dry friction between the plate and grains.
If there is no dry friction,
we confirmed that the perfect fluid model with the separation of the flow can reproduce the quantitative behavior of the drag force 
for the moderate dense case such as $\phi=0.72$, $0.74$, and $0.76$.
If there exists the dry friction, the yield stress $F_0(\phi)$ appears
and the perfect fluidity is no longer valid.
In this case, the drag force and the yield force diverge at the jamming point,
whose behavior is qualitatively similar to that observed in an experiment \cite{Takehara:2014}, 
but quantitative agreement between our results and their results is poor.
\section{Acknowledgement}
We would like to thank fruitful discussions 
with R.\ Blumenfeld, J.\ D.\ Goddard, K.\ Okumura, M.\ Otsuki and K.\ Suzuki.
We are also grateful to K.\ Saitoh for providing us the prototype of the program for the DEM simulation.
This work is partially supported by the Grant-in-Aid of MEXT (Grant No. 25287098).
%
%
%
%
\appendix\label{section:references}
%
%
\bibliography{paper}
%
%
%
\section{Notation}
\emph{The following symbols are used in this paper:}
\nopagebreak
\par
\begin{tabular}{r  @{\hspace{1em}=\hspace{1em}}  l}
$d$				& diameter of smaller grains;\\
$d_0$			& diameter of the tracer, equivalent to $D$;\\
$d_i$			& diameter of $i$-th particle;\\
$d_{ij}$			& average diameter of $i$-th and $j$-th particles;\\
$D$                    & diameter of the tracer; \\
$\bm{f}_{ij}^n$		& interaction between $i$-th and $j$-th particles in the normal direction;\\
$\bm{f}_{ij}^t$		& interaction between $i$-th and $j$-th particles in the tangential direction;\\
$\tilde{\bm{f}}_{ij}^t$& tangential interaction in the case of nonslip;\\
$F_0$			& yield force independent of the speed $V$;\\
$F_{\rm drag}$	& force exerted on the tracer;\\
$\bm{F}_{\rm ex}$	& external force exerted on the tracer;\\
$F_{\rm ex}$		& magnitude of $\bm{F}_{\rm ex}$;\\
$\bm{F}_{\rm int}$	& interaction among the tracer and disks;\\
$g$				& gravitational acceleration;\\
$I_i$				& moment of inertia of $i$-th particle;\\
$k_n$			& spring constant in the normal direction;\\
$k_t$			& spring constant in the tangential direction;\\
$m_i$			& mass of $i$-th particle; \\
$M$				& mass of the tracer; \\
$N$				& number of surrounding particles; \\
$r$				& radial coordinate in the polar coordinate;\\
$\bm{r}_0$		& position vector of the tracer, equivalent to $\bm{R}$;\\
$\bm{r}_i$		& position vector of $i$-th particle;\\
$r_{ij}$			& distance between $i$-th and $j$-th particles;\\
$\hat{\bm{r}}_{ij}$	& unit vector parallel to relative position vector of $i$-th and $j$-th particles;\\
\end{tabular}
\begin{tabular}{r  @{\hspace{1em}=\hspace{1em}}  l}
$\dot{\bm{r}}_{ij}$	& relative velocity of $i$-th and $j$-th particles;\\
$\bm{R}$			& position vector of the tracer;\\
$t$				& time;\\
$t_0$			& time of first contact of the particles;\\
$T_i$			& torque of $i$-th particle;\\
$\bm{t}_{ij}$		& unit vector parallel to the tangential force between $i$-th and $j$-th particles;\\
$u_r$			& radial component of the velocity near the tracer;\\
$u_r^*$			& radial component of the dimensionless velocity near the tracer;\\
$u_\theta$		& polar component of the velocity near the tracer;\\
$u_\theta^*$		& polar component of the dimensionless velocity near the tracer;\\
$\hat{\bm{v}}_i$	& unit vector parallel to the velocity of $i$-th particle;\\
$\bm{v}_{t,ij}$		& relative velocity between $i$-th and $j$-th particles in the tangential direction;\\
$V$				& steady speed of the tracer;\\
$\bm{V}$			& velocity of the tracer;\\
$V^*$			& dimensionless steady speed of the tracer;\\
$\hat{\bm{V}}$	& unit vector parallel to the velocity of the tracer;\\
$\alpha$			& coefficient of the term proportional to the square of the speed $V$;\\
$\beta$			& critical exponent of $\alpha$ with respect to the area fraction; \\
$\beta^\prime$	& critical exponent of the yield force $F_0$ with respect to the area fraction; \\
$\zeta_\omega$	& rolling friction constant;\\
$\eta_n$			& viscous parameter in the normal direction;\\
$\eta_t$			& viscous parameter in the tangential direction;\\
$\theta$			& angular coordinate in the polar coordinate;\\
$\theta_0$ 		& separation angle behind the tracer;\\
$\Theta(x)$		& The step function $\Theta(x)=1$ and $0$ for $x>0$ and $x\le0$, respectively;\\
$\mu$			& Coulombic friction constant between the tracer and the bottom plate;\\
$\mu_s$			& Coulombic friction constant between grains;\\
$\rho$			& mass density of granular fluid;\\
$\bm{\xi}_{ij}$		& tangential overlap vector between $i$-th and $j$-th particles;\\
$\phi$			& area fraction of the system; \\
$\phi_{\rm c}$		& jamming area fraction;\\
$\bm{\omega}_i$	& angular velocity of $i$-th particle; and\\
$\dot{\bm{\omega}}_i$	& angular acceleration of $i$-th particle.
\end{tabular}
\end{document}